\documentstyle[preprint,aps]{revtex}
\begin{document}

\draft

\newcommand{\lsim}{{ _< \atop ^\sim}}
\newcommand{\gsim}{{ _> \atop ^\sim}}
\setlength{\unitlength}{1mm}

\title{Thermodynamic Properties of the Incommensurate
Phase of $\bf CuGeO_3$}

\author{T. Lorenz, U. Ammerahl, R. Ziemes, and B. B\"uchner}
\address{II. Physikalisches Institut, Universit\"at zu K\"oln,
         Z\"ulpicher Str. 77, 50937 K\"oln, Germany}

\author{A.Revcolevschi, G. Dhalenne}
\address{Laboratoire de Chimie des Solides, Universit\'e Paris-Sud,
         91405 Orsay C\'edex, France}

\date{\today}

\maketitle

\begin{abstract}
We present high resolution measurements of
the specific heat and the thermal expansion
of the inorganic spin--Peierls cuprate $\rm CuGeO_3$
in a magnetic field of $\rm 16~Tesla$.
At the transition from the incommensurate to the uniform phase both
quantities show pronounced anomalies, which allow to derive the
uniaxial pressure dependencies of the transition temperature.
In high magnetic fields the specific heat is dominated by magnetic
excitations and follows a $\rm T^3$ law at low temperatures.
The thermal expansion measurements show the
occurrence of spontaneous strains along all three lattice constants
and yield high resolution measurements of the temperature
dependence of the incommensurate structural distortion.
The sizes of the spontaneous strains in the incommensurate phase
are significantly reduced, but both their anisotropy
as well as their temperature dependencies are
very similar to those in zero field.

\end{abstract}

\maketitle

During the last three years $\rm CuGeO_3$ has been the subject of intensive
investigations and the occurrence of a spin--Peierls transition ($\rm
T_{SP}\simeq 14 K$)
in this inorganic compound is well established now. Most of
the characteristic features of this transition are observed, e.g.
the opening of a gap in the magnetic excitation spectrum below
$\rm T_{SP}$~\cite{hase93a,nishi94b,fujita95a}
and the dimerization of the one--dimensional spin--$\rm
\frac{1}{2}$--Heisenberg chains~\cite{pouget94,hirota94,harris94}.
In addition to the dimerization,
very pronounced spontaneous strains ($\rm \epsilon$) of all three
lattice constants are found, which
are proportional to the square of
the structural order parameter Q~\cite{harris94,winkelmann95}.
This proportionality ($\rm \epsilon = kQ^2$) is
expected from a Ginzburg--Landau expansion of the free energy
when a usual linear quadratic coupling between Q and $\epsilon$
is assumed (see e.g. refs.~\cite{harris94,winkelmann95}).
The temperature magnetic field phase diagram~\cite{hase93c,poirier95a}
is also found to be in fair
agreement with the theoretical expectations~\cite{bulaevskii78,cross79b}.
For low fields a reduction of $\rm T_{SP}$
occurs which is proportional to $\rm H^2$.
For higher fields ($\rm H\gsim 12 T$) a transition to an incommensurate
phase takes place. In this I phase the lattice distortion is expected to be
incommensurate with respect to the underlying lattice.
There are, however, different models for
the spatial character of the incommensurate modulation;
domain--walls~\cite{nakano80}
or a sinusoidal modulation~\cite{cross79b}.
An incommensurate lattice modulation
is indeed observed in $\rm CuGeO_3$  for $\rm H \gsim 12 T$ by x--ray
scattering~\cite{kiryukhin95b,kiryukhin96a}. Although these measurements
favor the domain--wall picture, significant discrepancies between
theory and experiment are still present~\cite{kiryukhin96a}.

In order to study the magnetic and structural
properties of the I phase we have performed
high resolution measurements of such thermodynamic properties
as specific heat and thermal expansion.
Both quantities were measured on the same single crystal
of $\rm CuGeO_3$ with a size of about $\rm 6\times 5\times 8.3~mm^3$.
The specific heat ($\rm C_p$) was determined by a quasiadiabatic
step--by--step method.
The longitudinal thermal expansion coefficients
$\rm \alpha_i = (1/L_i)\partial L_i/\partial T$
($\rm L_i$ denotes the length of the sample
along the axis i) were measured with a capacitance dilatometer.

Fig.~\ref{cp} displays the specific heat data of $\rm CuGeO_3$ in a magnetic
field of 16~Tesla which was oriented along the $b$--axis.
For comparison, we also show the data relative to zero magnetic
field (see also refs.~\cite{sahling94,liu95b}).
It is obvious that
the main influence of the field is a strong decrease of both the transition
temperature and the size of the anomaly. Nevertheless, a sharp and rather large
anomaly of $\rm C_p$ is found at the transition between the I and the high
temperature uniform (U) phases.
The $\lambda$--like shapes of the anomalies indicate that in $\rm
CuGeO_3$ the transitions are strongly affected by fluctuations,
whereas the specific heat anomalies of the
organic spin--Peierls compounds seem to show
''mean--field'' behavior~\cite{wei76,northby82a,korving87},
i.e the ''upturns'' of $\rm C_p$ close to $\rm T_{SP}$ are
not observed. There are two possibilities to discuss
this difference of the anomaly shapes.
Either the better sample homogeneity of $\rm CuGeO_3$ allows to
observe the fluctuations at the spin--Peierls transition
or fluctuations are more pronounced in $\rm CuGeO_3$.
For example, one may speculate,
that the good agreement with mean--field
behavior in the organic compounds
is related to the pre--existing soft phonon
which has not been found in $\rm CuGeO_3$ so far.

Although the broadening of the transitions associated with sample
inhomogeneities is rather small in $\rm CuGeO_3$, there is a significant
asymmetry with regard to the maximum of $\rm C_p(T)$.
Due to this asymmetry a description
of the anomaly shape by a 'purely' critical behavior of $\rm C_p$ is
questionable. As shown by the authors of ref.~\cite{liu95b} for their
zero field data, which
-- although obtained on a different crystal -- are in very good agreement
with our data,  it is not possible
to unambiguously assign a universality class to the transition via the
critical exponent of $\rm C_p$.
We note that -- apart from their different
sizes -- the shapes of the anomalies close to $\rm T_{SP}$ do not
differ significantly in $\rm H=0$ and $\rm 16 T$. Thus, we do not
analyze the critical behavior of $\rm C_p$ in this communication.

In the low temperature range, however, we find distinct differences
between the zero and high field data. The Inset of Fig.~\ref{cp}
shows $\rm C_p/T$
versus $\rm T^2$ for H=0 and 16 Tesla, respectively. As expected the data in
zero field exhibit a clear curvature down to
the lowest temperatures in this representation. This is due to the
activated behavior of the magnetic
contribution of $\rm C_p$. The solid line represents a fit in terms of
$\rm C_p=\beta_{ph} T^3+\delta\cdot exp(-\Delta E/T)$. Restricting
the fit to $\rm T<6~K$ we obtain a value
$\rm \beta_{ph} = 0.3~mJ~mole^{-1}~K^{-4}$ for the lattice contribution.
The magnetic contribution is given by
$\rm \delta = 3.6 ~J~mole^{-1}~K^{-1}$ and the energy gap amounts
to $\rm \Delta E = 23~K$,
in fair agreement with the results reported in ref.~\cite{liu95b}.

The high field data show a completely different low temperature behavior.
The entire specific heat follows a pure $\rm T^3$ law.
There is no indication for a gap in the magnetic excitation spectrum of the I
phase. Assuming that $\rm \beta_{ph}$ does not depend on H
the specific heat is described by
$\rm C_p=(\beta_{ph} + \beta_{mag})\cdot T^3$.
For $\rm \beta_{mag}$
we determine a value of $\rm 1.4~mJ~mole^{-1}~K^{-4}$,
which is significantly larger than $\rm \beta_{ph}$. Thus,
in the low temperature range the specific
heat of the I phase is dominated by the magnetic excitations, which are
strongly suppressed in zero field due to the opening of the gap.

The $\rm T^3$ dependence of the magnetic specific heat
is present in a rather large temperature range in the I
phase, i.e. up to 5.5 K $\rm \gsim T_{SP}/2$. Moreover,
we also find the  $\rm T^3$ law with the {\em same} coefficient $\rm \beta_{mag}$
for smaller magnetic fields down to 12.5 Tesla.
Apparently, neither the temperature dependence nor the
absolute value of $\rm C_p$ strongly changes as a function of the magnetic
field in the I phase. A very strong field dependence of
both the temperature dependence and the absolute value of $\rm C_p$
at low temperatures is, however, present in the D--phase
as can be extracted already from a comparison
of the data in 0 T and 6 T shown in ref.~\cite{liu95b}.

Let us shortly discuss the implications of
the $\rm T^3$ law we observe for the magnetic specific heat
of the I phase.
The -- to our knowledge -- only existing theoretical calculation of $\rm C_p$,
which uses the soliton lattice solution
for the I phase based on a mean--field Hamiltonian~\cite{fujita84},
can not explain our experimental finding. Fujita and Machida
find in their calculations a BCS behavior of $\rm C_p$ for both
zero field as well as in the I phase, i.e. they
obtain a finite but reduced gap in the I phase as well.
Even if we assume a very small value of this gap,
it is impossible to fit our data with the
theoretically expected BCS temperature dependence.

From the experimentally observed
specific heat one can in principle extract the dispersion relations
of the magnetic excitations. Assuming a Bose--statistics suggested
by the $\rm T^3$ law and a dispersion relation of the form
$\rm \omega \propto k^n$ the specific heat at low temperatures
is proportional to $\rm T^{D/n}$ where D denotes the dimension
of the system. For three--dimensional excitations
our finding implies linear dispersion relations
in the I phase of $\rm CuGeO_3$ as in usual antiferromagnets.
For D = 1 (or 2) rather strange dispersion
relations are obtained, which makes the assumption
of low--dimensional magnetic excitations in the I phase
rather unlikely. Further theoretical and experimental work is necessary
to decide, whether there are really three dimensional
excitations with $\rm \omega\propto k$ in the I phase or whether
the $\rm T^3$ law of $\rm C_p$ we observe in a large
temperature and field range is due to an accidental
superposition of different contributions.

Our findings for the thermal expansion in the incommensurate phase
markedly differ from those of the specific heat.
In Fig.~\ref{al} we show the thermal expansion coefficients
along the $a$, $b$, and
$c$ axes. Concerning the thermal expansion one has to distinguish between the
structural and magnetic anisotropy. The magnetic anisotropy arises
from the different values of the gyromagnetic ratios~\cite{ohta94}
along the different crystal axes.
In our experimental setup
the magnetic field is always parallel to the crystal axis
whose thermal expansion coefficient is measured.
Therefore, the transition temperatures slightly differ in the
measurements carried out along the different crystal
axes (see Fig.~\ref{al} and Table~\ref{vonp}).

The thermal expansion data in $\rm H=0$ are also shown
in Fig.~\ref{al}. For each lattice constant the zero and high
field data were obtained during the same run, i.e. for exactly the same
orientation of the crystal.
The anomalies occurring at the I/U transition are strongly reduced in size
(by about a factor 2) compared to those found in zero field.
Despite this strong decrease, the anomalies are still
rather large and indicate that the U/I transition strongly
depends on pressure (see below). Furthermore, the anomalies reveal the
occurrence of spontaneous strains within the I phase, which
are also seen in the temperature
dependencies of the lattice constants (Fig.~\ref{eps}).
The spontaneous strains are obtained by integration of the
anomalous contribution $\rm \delta\alpha_i\equiv\alpha_i-\alpha_{i,extr.}$
below $\rm T_{SP}$. The $\rm \alpha_{i,extr.}$
are smooth polynomials (given by the dashed lines in
Fig.~\ref{al}) representing the
extrapolation of the $\rm \alpha_i$ in the U phase to 0 at T = 0
K~\cite{winkelmann95}.

In zero field the $\rm \epsilon_i$ are proportional to the square
of the order parameter~\cite{harris94,winkelmann95,we_tad}.
The characteristic temperature dependence
of an order parameter is also visible in
H = 16 T (right part of Fig.~\ref{eps}). The
temperature dependence of the structural order parameter in the I phase,
i.e. the incommensurate lattice deformation, has not been measured
by diffraction techniques so far. Thus, we can not unambiguously
prove $\rm \epsilon_i\propto Q^2$ in the I phase.
However, this leading order of the strain order parameter
coupling is usually observed at
structural transitions and, moreover, there is
no indication that higher order terms are important
at the U/I transition in $\rm CuGeO_3$.
In the following we will assume $\rm \epsilon\propto Q^2$ for
both phases and a comparison
of the $\rm \epsilon_i$ in $\rm H=16T$ and $\rm H=0$ allows
for a comparison of the structural order parameters in the
I and D phases, respectively.

Besides differences in the critical behavior of the order parameter
close to $\rm T_{SP}$, which will be discussed elsewhere~\cite{we_tad},
the $\rm \epsilon_i$ occurring at the I phase compare well to those
in zero field.
In Fig.~\ref{alpred}
we show $\rm T_{SP}(H)\cdot \delta\alpha_i=
- \frac{\partial \epsilon_i}{\partial t}$ versus reduced temperature
$\rm t=1-T/T_{SP}(H)$.
That means, we compare the derivatives
$\rm \frac{\partial \epsilon_i}{\partial t}
= k_i^{D/I} \frac{\partial Q^2}{\partial t}$ of the spontaneous
strains occurring in the D and I phase, respectively.
($\rm k_i^{D/I}$ denote the strain order parameter coupling constants.)
Please note that by comparing the temperature derivatives of the
$\rm \epsilon_i(t)$ possible differences of their temperature
dependencies in $\rm H=0$ and $\rm H=16~Tesla$ will show up more
clearly than by comparing the $\rm \epsilon_i(t)$ themselves.
It is apparent from Fig.~\ref{alpred} that, apart from
the absolute values (see the different scales in Fig.~\ref{alpred}),
the temperature derivatives of the $\rm \epsilon_i$ in the I phase
are very similar to those in $\rm H=0$. This holds for all three crystal axes.
Moreover, the ratio between
$\rm \frac{\partial \epsilon_i}{\partial t}$ in $\rm H=0$ and in $\rm H=16~T$
is independent on the crystal axis and
amounts to a value of 3, i.e. the structural anisotropy does not
change in a magnetic field.
We emphasize that for the specific heat in $\rm H=0$ and $\rm H=16T$
such a similarity is not present
at all (see Fig.~\ref{cp}). The magnetic and the structural
degrees of freedom show completely different magnetic field dependencies.

The reduced $\rm \epsilon_i$ in the I phase
may arise either from three wave vector dependent strain order parameter
coupling constants ($\rm k_i^I=k_i^I(q)$) or a reduced average amplitude
A of the distortion. The identical anisotropy in $\rm H=0$ and $\rm H=16T$
strongly indicates that the reduction for all three lattice constants is
determined by a common parameter, e.g. by A.
There is a straightforward qualitative explanation of such
a reduction within the domain wall picture.
A is expected to reduce when approaching the domain walls\cite{nakano80}.
The reduced $\rm \epsilon_i$ in the I phase is then a consequence of
the reduced average value of the dimerization within a single
domain. Moreover, the $\rm \epsilon_i$ should show a further decrease
with increasing field due to the increasing number of domains.
However, a sinusoidal modulation of the lattice distortion
($\rm A(x)=A_0 sin(qx)$) also
causes a reduction of the average value of the dimerization.
Thus, from our present data it is impossible
to discriminate between these two models of the structural distortion.
Measurements up to still higher fields are planned to study the field
dependence of
$\rm \epsilon_i(T,H)$ in the I phase to clarify this question.

Finally we derive the uniaxial pressure dependencies of the U/I
transition by comparing the anomaly sizes of the
$\rm \alpha_i$ and $\rm C_p$~\cite{cpvonh} as described
in ref.~\cite{winkelmann95,we_tad}.
Similar as for H = 0 T
the uniaxial pressure dependencies of $\rm T_{SP}$ in $\rm H=16T$
are very large and strongly anisotropic (see Table~\ref{vonp}).
For pressure along the $a$ axis $\rm T_{SP}$ decreases,
whereas it increases for pressure along the two other axes.
The hydrostatic pressure dependence given by the sum of the
$\rm \partial T_{SP}/\partial p_i$
amounts to $\rm +3.6 K/GPa$. All the
pressure dependencies in $\rm H=16T$
are significantly smaller than those in $\rm H=0$ (Table~\ref{vonp}).
There are, however, pronounced similarities to the zero field data.
The anisotropies of the $\rm \partial T_{SP}/\partial p_i$
at the U/I transition and at $\rm H=0$
are nearly identical. Moreover, the pressure induced
relative changes of $\rm T_{SP}$
in $\rm H=0$ and $\rm H=16T$ are very similar.
For instance, along the $b$ axis an increase of
about 50\% /GPa is found for both
$\rm T_{SP}(H=0)$ and $\rm T_{SP}(H=16T)$ indicating
a pressure independent H/T phase diagram
in reduced scales, which is in
agreement with theoretical predictions.
However, as we have recently shown~\cite{buechner96a}
the $\rm \partial T_{SP}/\partial p_i$ unexpectedly
correlate with the magnetoelastic coupling.

To summarize, we have reported measurements of the specific heat and
the thermal expansion of $\rm CuGeO_3$ in $\rm H=16T$. For both
quantities we find pronounced anomalies at the U/I transition.
The specific heat at low temperatures is dominated by the
magnetic excitations.
In contrast to the findings in zero field
our data in $\rm H=16T$ yield no indication
for a gap in the magnetic excitations.
Instead the magnetic specific heat follows
a pure $\rm T^3$ law.
The thermal expansion below $\rm T_{SP}$ is dominated by the
incommensurate lattice distortion leading
to spontaneous strains of all three lattice constants which scale
with the structural order parameter.
Remarkably, the temperature dependencies of the
commensurate and incommensurate lattice distortion
at H = 0 and 16 T, respectively, are very similar,
whereas the magnetic specific heats markedly differ.

The work at Cologne was supported by the Deutsche Forschungsgemeinschaft
through SFB 341. U.A. acknowledges support by the Graduiertenkolleg
GRK14 of the Deutsche Forschungsgemeinschaft.

\begin{figure}       
\caption[]{Specific heat of $\rm CuGeO_3$
in H=0 ($\circ$) and 16 Tesla ($\bullet$). Inset:
$\rm C_p/T$ vs. $\rm T^2$; the solid lines are fits
assuming activated behavior ($\rm \circ,~H=0$) and a $\rm T^3$ law
($\rm \bullet,~H=16T$, see text), respectively.}
\label{cp}
\end{figure}

\begin{figure}       
\caption[]{Thermal expansion coefficients of $\rm CuGeO_3$
in H=0 ($\bullet$) and 16 Tesla ($\circ$). The dashed lines
represent the extrapolated behavior of $\rm \alpha_i$ of the U phase
to $\rm \alpha_i(T=0)=0$.
The arrows mark the transition temperatures derived at the maximum slopes
of $\rm \alpha_i (T)$}
\label{al}
\end{figure}

\begin{figure}       
\caption[]{Left panel: Temperature dependence of the lattice constants
of $\rm CuGeO_3$ in H=16 Tesla.
Right panel: Spontaneous strains $\rm \epsilon_i=\int
\alpha_{i}-\delta\alpha_{i,extr.}$ in the I phase.}
\label{eps}
\end{figure}

\begin{figure}       
\caption[]{Temperature derivatives of the spontaneous strains
$\rm -\frac{\partial \epsilon_i}{\partial t}
\propto \frac{\partial Q^2}{\partial t} $ vs. reduced temperature
t in $\rm H=16T$ ($\rm \circ$, left scale) and
$\rm H=0$ ($\rm \bullet$, right scale).}
\label{alpred}
\end{figure}

\begin{table}
\begin{tabular}{c|c|cc}
axis   & H=0  $\rm T_{SP}$=14.35(10)K  & \hfill H = &\hskip-6cm $\rm 16~Tesla$ \\
\hline
   &   $\rm \frac{\partial T_{SP}}{\partial p_i}\mid_{p_i=0}$  (K/GPa) &
   $\rm T_{SP}$  (K)      &
                      $\rm \frac{\partial T_{SP}}{\partial p_i}\mid_{p_i=0}
   (K/GPa)$ \\
\hline
 $a$ &  -3.7(5) & 10.13(10)   &  -2.8(5) \\
 $b$ &   7.2(5) & 10.02(10)   &   5.3(5)  \\
 $c$ &   1.6(5) & 10.23(10)   &   1.1(5)
\end{tabular}

\caption[]{Uniaxial pressure dependencies of $\rm T_{SP}$ along
the different crystal
axes in $\rm H=0$ and $\rm H=16T$ as calculated from
the thermal expansion and specific heat anomalies~\cite{cpvonh}.
The different $\rm T_{SP}$'s in
$\rm H=16T$ are due to the slightly different g--factors of $\rm CuGeO_3.$
\label{vonp}}
\end{table}


\begin{thebibliography}{10}
\bibitem{we_tad}
A detailed discussion of the measurements of the thermal expansion in the
  entire field range between 0 and 16 Tesla will be given in a forthcoming
  publication.
\bibitem{cpvonh}
We confirmed this assumption by measurements of $\rm C_p$ in magnetic fields
  applied along a different direction.
\end{thebibliography}
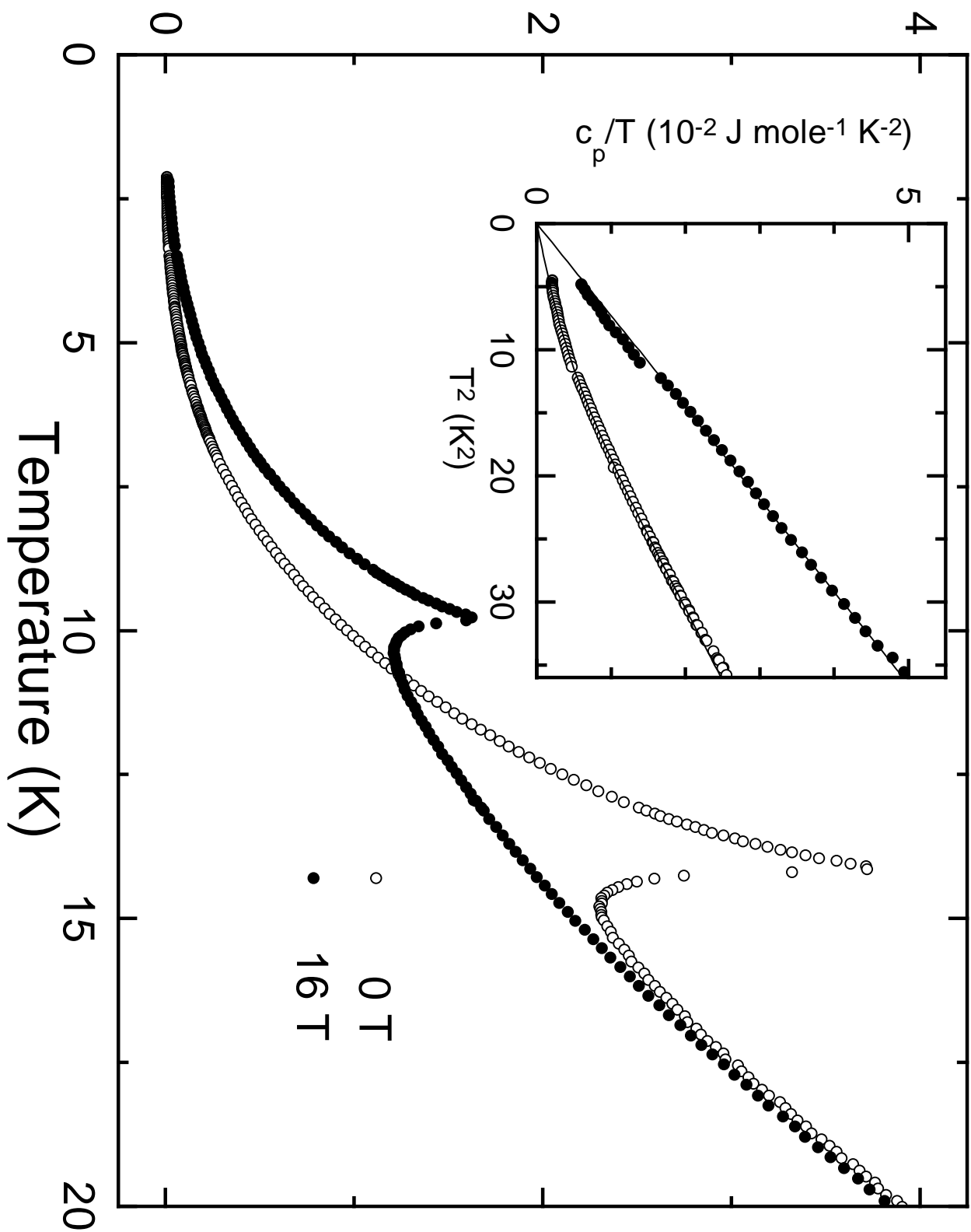
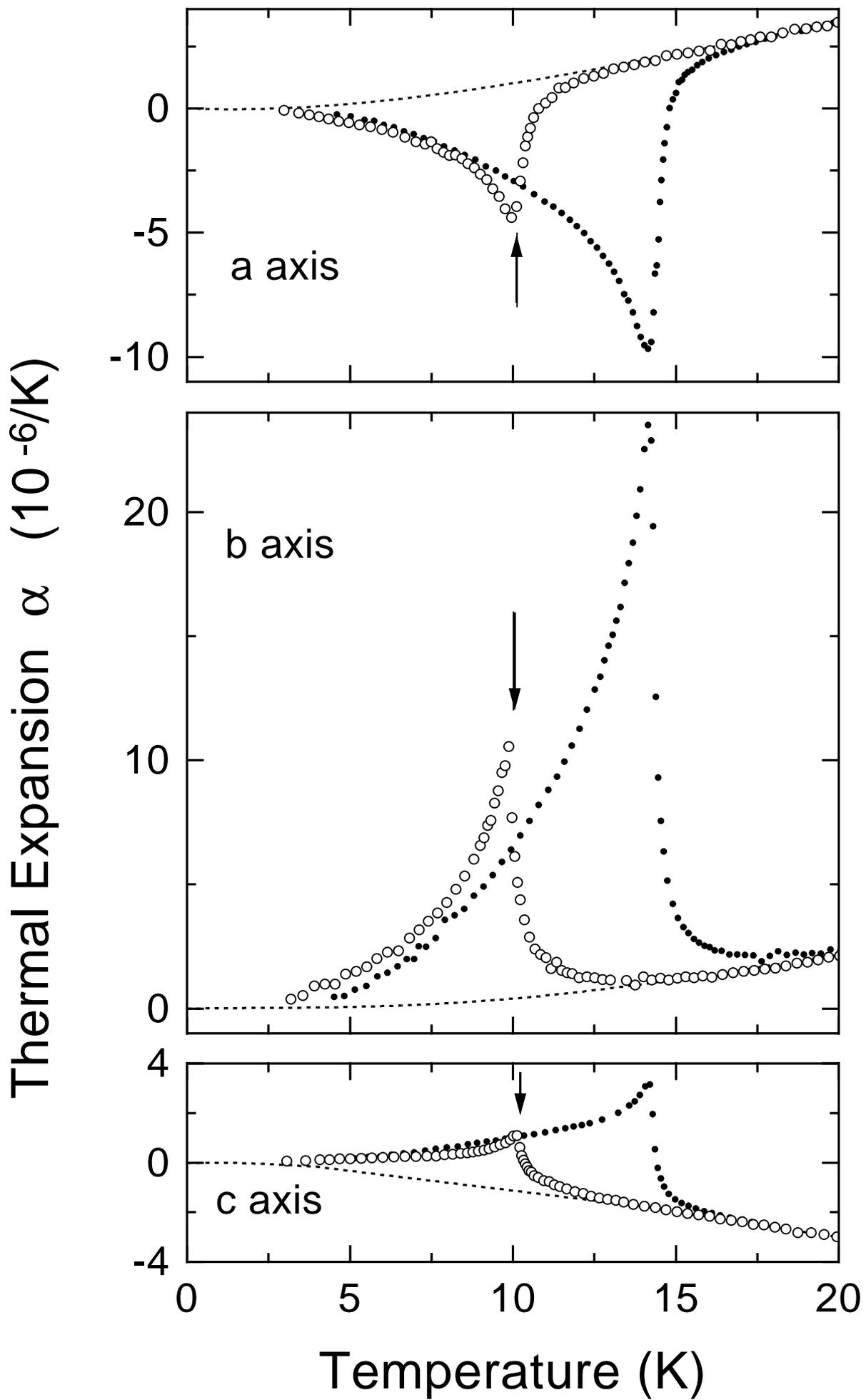
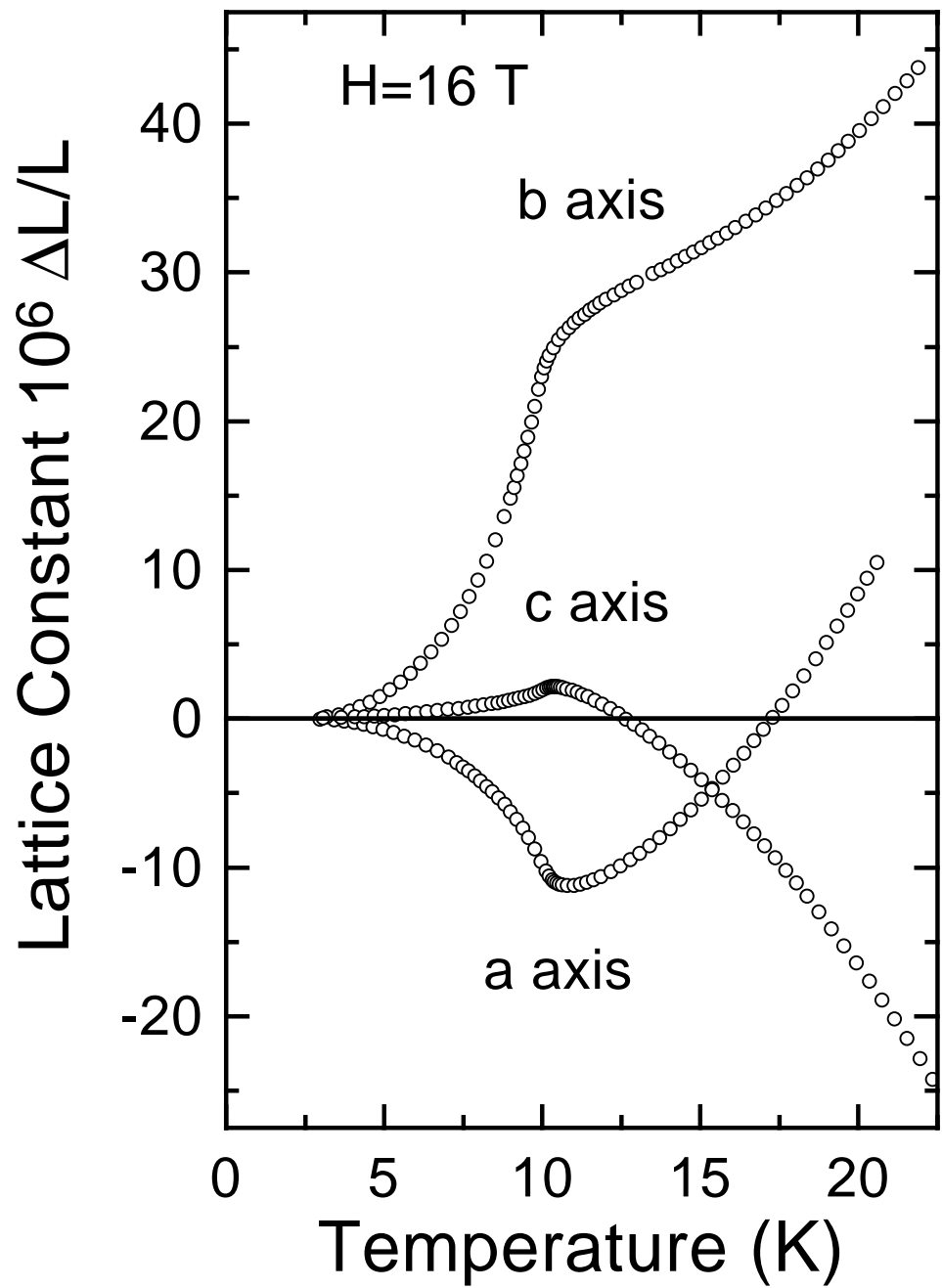
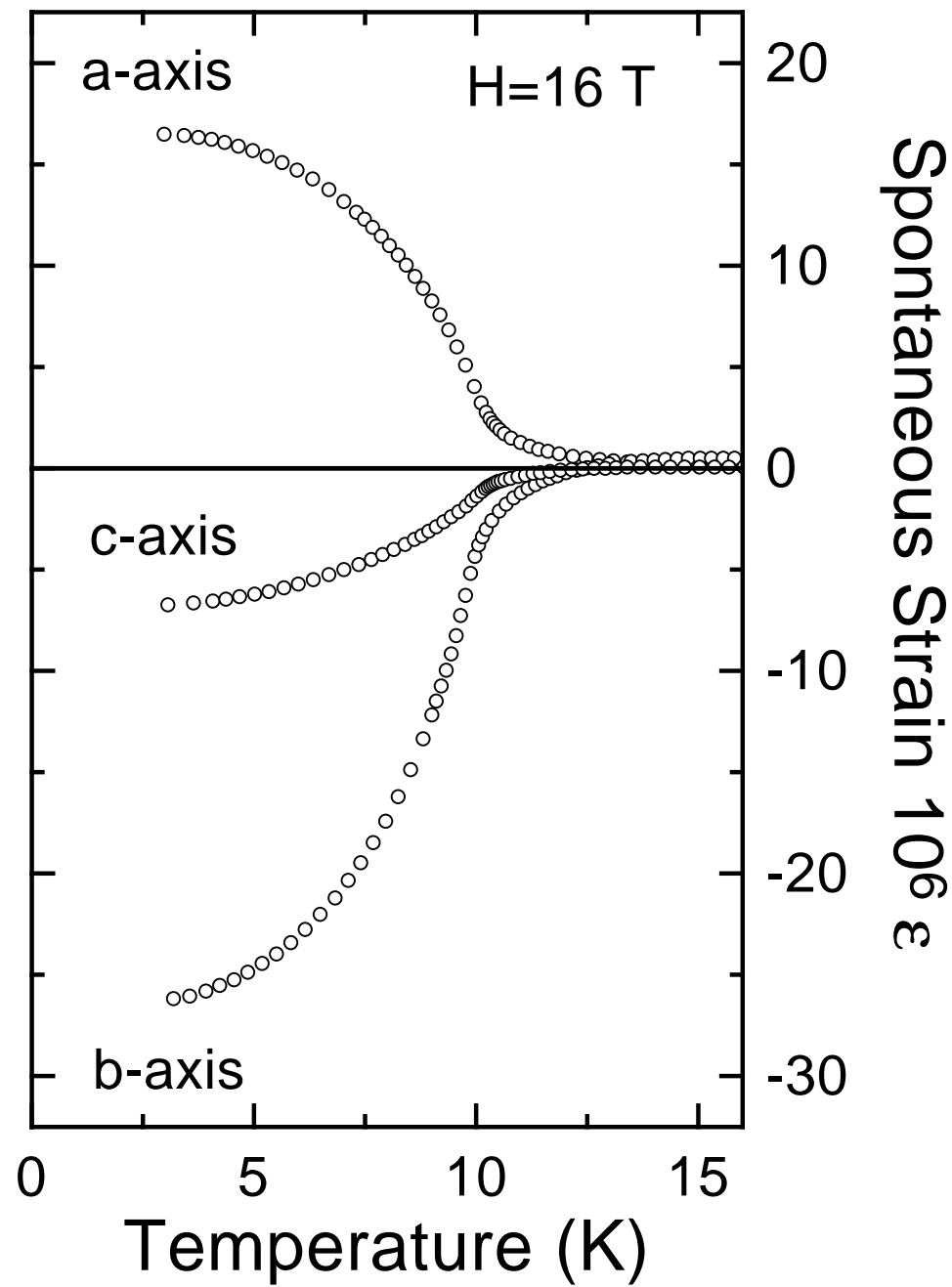
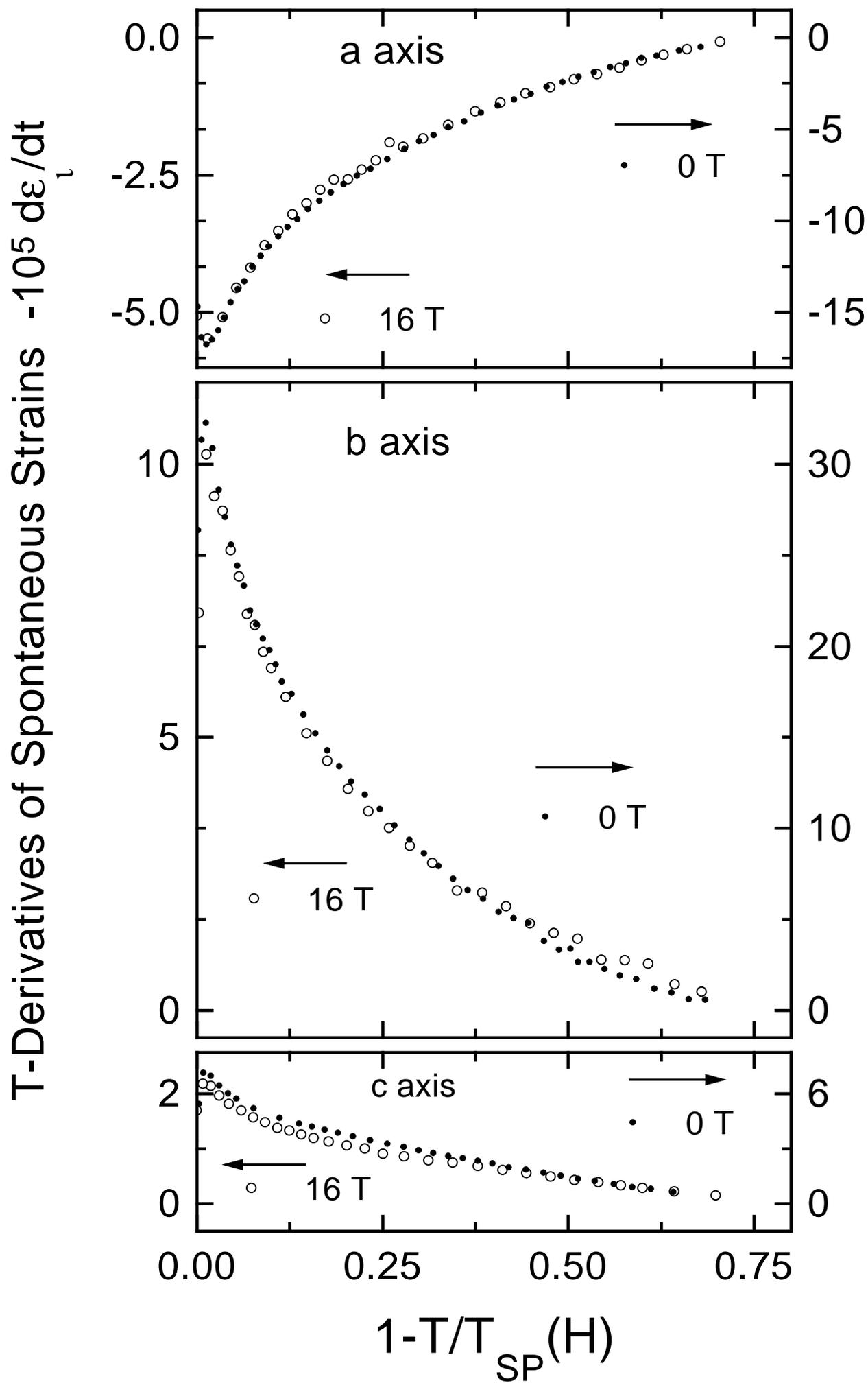
\end{document}